\documentclass[twocolumn,english]{revtex4-2}
\usepackage[utf8]{inputenc}
\usepackage[T1]{fontenc}
\usepackage{amsmath}
\usepackage{amsfonts}
\usepackage{amssymb}
\usepackage[version=4]{mhchem}
\usepackage{stmaryrd}
\usepackage{hyperref}
\hypersetup{colorlinks=true, linkcolor=blue, filecolor=magenta, urlcolor=cyan,}
\usepackage{mathrsfs}

\begin{document}
\title{Anti-Jahn-Teller disproportionation and prospects for spin-triplet superconductivity in $d$-element compounds }

\author{A.S. Moskvin, Yu.D. Panov}

\affiliation{Ural Federal University, Yekaterinburg, Russia}

\begin{abstract}
We argue that the unusual properties of a wide class of materials based on Jahn-Teller $3 d$ - and $4 d$-ions with different crystal and electronic structures, from quasi-two-dimensional unconventional superconductors (cuprates, nickelates, ferropnictides/chalcogenides, ruthenate $\mathrm{Sr}_{2} \mathrm{RuO}_{4}$ ), manganites with local superconductivity to 3D ferrates $(\mathrm{CaSr}) \mathrm{FeO}_{3}$, nickelates $\mathrm{RNiO}_{3}$ and silver oxide $\mathrm{AgO}$ with unusual charge and magnetic order can be explained within a single scenario. The properties of these materials are related to the instability of their highly symmetric Jahn-Teller "progenitors" with the ground orbital E-state to charge transfer with anti-Jahn-Teller disproportionation and the formation of a system of effective local composite spin-singlet or spin-triplet, electronic or hole bosons moving in a non-magnetic or magnetic lattice. These unusual systems are characterized by an extremely rich variety of phase states from non-magnetic and magnetic insulators to unusual metallic and superconducting states.
\end{abstract}

\keywords{Jahn-Teller effect, disproportionation, local composite bosons, spin-triplet superconductivity.}

\maketitle

\section{Introduction}
Investigation of spin-triplet superconductivity (STS) is of fundamental research and practical importance most of all because, on the one hand, formations of $S=1$ superconducting carriers cannot be by any means explained within the traditional Bardeen-Cooper-Schrieffer (BCS) theory, and, on the other hand, bound Majorana states may be implemented in the spin-triplet superconductors to be further used in topological quantum computing. However, currently there is neither a single STS theory or solid-state materials with reliably ascertained STS. Even in strontium ruthenate $\mathrm{Sr}_{2} \mathrm{RuO}_{4}$, which has been considered as an example of the first STS material for many years, the spin nature of superconducting carriers remains a matter of heated discussions (see the review [1]). Earlier in [2], one of the authors offered a mechanism for spin-triplet composite boson formation as a result of so-called ``anti-Jahn-Teller'' disproportionation in compounds based on nominally JahnTeller (JT) 3d-ions or JT magnets. This mechanism pointed out, in particular, spin-triplet superconductivity of Fe-pnictides and Fe-chalcogenides predicted as early as in 2008 [3]. Over the past years, new research results for JT magnets based on both $3 d$ - and $4 d$-ions, first of all ruthenates, as well as new arguments both for and against spin-triplet superconductivity have been obtained. This study expands the ``anti-Jahn-Teller'' disproportionation model to a wider class of JT magnets than in [2], including $4 d$-magnets (ruthenates, silver compounds) and 2D nickelates $\mathrm{RNiO}_{2}$ and shows that all they may be described within a single scenario. Section II describes a class of JT magnets and their evolution resulting from the competition of the Jahn-Teller stabilization energy and $d-d$-disproportionation. Section III addresses possible phase states of JT magnets. Section IV summarizes effective Hamiltonians of the spin-triplet boson system in non-magnetic and magnetic lattices. A concise conclusion is provided in Section V.

\section{Jahn-Teller magnets}
Jahn-Teller (JT) magnets include compounds based on Jahn-Teller $3 d$ - and $4 d$-ions with $t_{2 g}^{n_{1}} e_{g}^{n_{2}}$ type configurations in highly symmetric octahedral, cubic or tetrahedral environment and with ground orbital E-doublet [2]. These are compounds based on tetra-complexes with $d^{1}$ configuration $\left(\mathrm{Ti}^{3+}, \mathrm{V}^{4+}\right)$ and high-spin (HS) $d^{6}$ configuration $\left(\mathrm{Fe}^{2+}, \mathrm{Co}^{3+}\right)$, octa-complexes with HS $d^{4}$ configuration $\left(\mathrm{Mn}^{3+}, \mathrm{Fe}^{4+}, \mathrm{Ru}^{4+}\right)$, low-spin (LS) $d^{7}$ configuration $\left(\mathrm{Co}^{2+}\right.$, $\left.\mathrm{Ni}^{3+}, \mathrm{Pd}^{3+}\right)$, and octa-complexes with $d^{9}$ configuration $\left(\mathrm{Cu}^{2+}, \mathrm{Ni}^{1+}, \mathrm{Ag}^{2+}\right)$ (see the Table I). The JT magnet class includes a many promising materials which are in the focus of the modern condensed matter physics such as manganites $\mathrm{RMnO}_{3}$, ferrates $(\mathrm{Ca}, \mathrm{Sr}) \mathrm{FeO}_{3}$, ruthenates $\mathrm{RuO}_{2},(\mathrm{Ca}, \mathrm{Sr}) \mathrm{RuO}_{3},(\mathrm{Ca}, \mathrm{Sr})_{2} \mathrm{RuO}_{4}$, wide set of Fe-pnictides $(\mathrm{FePn})$ and Fe-chalcogenides $(\mathrm{FeCh}), 3 \mathrm{D}$ nickelates $\mathrm{RNiO}_{3}$, $3 \mathrm{D}$ cuprate $\mathrm{KCuF}_{3}, 2 \mathrm{D}$ cuprates $\left(\mathrm{La}_{2} \mathrm{CuO}_{4}, \ldots\right)$ and nickelates $\mathrm{RNiO}_{2}$, and silver-based compounds $\left(\mathrm{AgO}, \mathrm{AgF}_{2}\right)$ (see the Table I). These materials have a wide range of unique properties from various types of magnetic and charge ordering to metal-insulator transition and superconductivity.

\begin{table*}[thb]
\caption{Jahn-Teller $3 d^{n}$ and $4 d^{n}$-systems which are optimal for formation of superconductivity induced by ``anti-JT'' disproportionation. The last column shows the examples of actual compounds}
{ 
\global\long\def\arraystretch{1.4}%
 
\begin{tabular}{|c|c|c|c|c|c|}
\hline
\begin{tabular}{l}
JT conf. \\
JT ions \\
\end{tabular} & Sym. & LS/HS & \begin{tabular}{l}
Local \\
boson \\
\end{tabular} & Lattice & \begin{tabular}{l}
Compound \\
examples \\
\end{tabular} \\
\hline
\begin{tabular}{c}
$3 d^{1}\left(e_{g}^{1}\right):{ }^{2} E$ \\
$\mathrm{Ti}^{3+}, \mathrm{V}^{4+}$ \\
\end{tabular} & tetra & - & \begin{tabular}{c}
$e_{g}^{2}:{ }^{3} A_{2 g}$ \\
$s=1$ \\
\end{tabular} & \begin{tabular}{c}
$A_{1 g}$ \\
$S=0$ \\
\end{tabular} & $?$ \\
\hline
\begin{tabular}{c}
$3 d^{4}\left(t_{2 g}^{3} e_{g}^{1}\right):{ }^{5} E$ \\
$\mathrm{Mn}^{3+}, \mathrm{Fe}^{4+}$ \\
\end{tabular} & octa & HS & \begin{tabular}{c}
$e_{g}^{2}:{ }^{3} A_{2 g}$ \\
$s=1$ \\
\end{tabular} & \begin{tabular}{c}
$A_{2 g}$ \\
$S=3 / 2$ \\
\end{tabular} & \begin{tabular}{c}
$(\mathrm{Ca}, \mathrm{Sr}) \mathrm{FeO}_{3}$ \\
$\mathrm{RMnO}_{3}$ \\
\end{tabular} \\
\hline
\begin{tabular}{c}
$4 d^{4}\left(t_{2 g}^{3} e_{g}^{1}\right):{ }^{5} E$ \\
$\mathrm{Ru}^{4+}$ \\
\end{tabular} & octa & HS & \begin{tabular}{c}
$e_{g}^{2}:{ }^{3} A_{2 g}$ \\
$s=1$ \\
\end{tabular} & \begin{tabular}{c}
$A_{2 g}$ \\
$S=3 / 2$ \\
\end{tabular} & \begin{tabular}{c}
$\mathrm{RuO}_{2}$ \\
$(\mathrm{Ca}, \mathrm{Sr})_{2} \mathrm{RuO}_{4}$ \\
$(\mathrm{Ca}, \mathrm{Sr}) \mathrm{RuO}_{3}$ \\
\end{tabular} \\
\hline
\begin{tabular}{c}
$3 d^{6}\left(e_{g}^{3} t_{2 g}^{3}\right):{ }^{5} E$ \\
$\mathrm{Fe}^{2+}, \mathrm{Co}^{3+}$ \\
\end{tabular} & tetra & HS & \begin{tabular}{c}
$\underline{e}_{g}^{2}:{ }^{3} A_{2 g}$ \\
$s=1$ \\
\end{tabular} & \begin{tabular}{c}
$A_{1 g}$ \\
$S=3 / 2$ \\
\end{tabular} & FePn, FeCh \\
\hline
\begin{tabular}{c}
$3 d^{7}\left(t_{2 g}^{6} e_{g}^{1}\right):{ }^{2} E$ \\
$\mathrm{Co}^{\mathrm{II}+}, \mathrm{Ni}^{\mathrm{iII}+}$ \\
\end{tabular} & octa & LS & \begin{tabular}{c}
$e_{g}^{2}:{ }^{3} A_{2 g}$ \\
$s=1$ \\
\end{tabular} & \begin{tabular}{c}
$A_{1 g}$ \\
$S=0$ \\
\end{tabular} & \begin{tabular}{c}
$\mathrm{RNiO}_{3}$ \\
$\mathrm{AgNiO}_{2}$ \\
\end{tabular} \\
\hline
\begin{tabular}{c}
$3 d^{9}\left(t_{2 g}^{6} e_{g}^{3}\right):{ }^{2} E$ \\
$\mathrm{Cu}^{2+}, \mathrm{Ni}^{+}$ \\
\end{tabular} & octa & - & \begin{tabular}{c}
$\underline{e}_{g}^{2}:{ }^{3} A_{2 g}$ \\
$s=1$ \\
\end{tabular} & \begin{tabular}{c}
$A_{1 g}$ \\
$S=0$ \\
\end{tabular} & \begin{tabular}{c}
$\mathrm{KCuF}_{3}$ \\
$\mathrm{~K}_{2} \mathrm{CuF}_{4}$ \\
\end{tabular} \\
\hline
\begin{tabular}{c}
$4 d^{9}\left(t_{2 g}^{6} e_{g}^{3}\right):{ }^{2} E$ \\
$\mathrm{Ag}^{2+}$ \\
\end{tabular} & octa & - & \begin{tabular}{c}
$\underline{e}_{g}^{2}:{ }^{3} A_{2 g}$ \\
$s=1$ \\
\end{tabular} & \begin{tabular}{c}
$A_{1 g}$ \\
$S=0$ \\
\end{tabular} & $\mathrm{AgO}$ \\
\hline
\begin{tabular}{c}
$3 d^{9}\left(t_{2 g}^{6} e_{g}^{3}\right):{ }^{2} B_{1 g}$ \\
$\mathrm{Cu}^{2+}, \mathrm{Ni}^{+}$ \\
\end{tabular} & \begin{tabular}{l}
octa $^{*}$ \\
square \\
\end{tabular} & - & \begin{tabular}{c}
$\underline{b}_{1 g}^{2}:{ }^{1} A_{1 g}$ \\
$s=0$ \\
\end{tabular} & \begin{tabular}{c}
$A_{1 g}$ \\
$S=0$ \\
\end{tabular} & \begin{tabular}{c}
HTSC cuprates \\
$\mathrm{CuO}, \mathrm{RNiO}_{2}$ \\
\end{tabular} \\
\hline
\begin{tabular}{c}
$4 d^{9}\left(t_{2 g}^{6} e_{g}^{3}\right):{ }^{2} B_{1 g}$ \\
$\mathrm{Ag}^{2+}$ \\
\end{tabular} & square & - & \begin{tabular}{c}
$\underline{b}_{1 g}^{2}:{ }^{1} A_{1 g}$ \\
$s=0$ \\
\end{tabular} & \begin{tabular}{c}
$A_{1 g}$ \\
$S=0$ \\
\end{tabular} & $\mathrm{AgF}_{2}$ \\
\hline
\end{tabular}

} 
\end{table*}


Lifting the orbital degeneracy in JT magnets may be due to the specific features of crystalline structure like,
for example, in ``apexless'' $2 \mathrm{D}$ cuprates $\left(\mathrm{Nd}_{2} \mathrm{CuO}_{4}\right)$ and nickelates $\mathrm{RNiO}_{2}$, and also to Jahn-Teller effect which usually results in formation of an antiferromagnetic insulator phase (see, for example, $\mathrm{La}_{2} \mathrm{CuO}_{4}, \mathrm{KCuF}_{3}$ ). A competing orbital degeneracy lifting mechanism in the JT magnets considered above is the ``anti-Jahn-Teller'' $d-d$ disproportionation according to scheme
\begin{equation}
	3 d^{n}+3 d^{n} \rightarrow 3 d^{n+1}+3 d^{n-1},
\end{equation}
involving formation of system of bound or relatively free $3 d^{n+1}$ electron and $3 d^{n-1}$ hole centers featuring an electron/hole pair. An electron/hole center can be formally represented as a hole/electron center with an electron/hole pair $3 d^{2} / \underline{3}^{2}$ localized in the center. In other words, the disproportionation system can be formally represented as a system of local spin-singlet or spin-triplet composite electron/hole bosons ``moving'' on the hole/electron center lattice. Note that the disproportionation energy formally coincides with the local correlation energy $U$.

In the systems with strong $d-p$-hybridization (cationanion covalence), the disproportionation reaction (1) shall be written in the ``cluster'' language, for example, like for $\mathrm{CuO}_{4}$ clusters in $\mathrm{CuO}_{2}$ planes of cuprates
\begin{equation}
	\left[\mathrm{CuO}_{4}\right]^{6-}+\left[\mathrm{CuO}_{4}\right]^{6-} \rightarrow\left[\mathrm{CuO}_{4}\right]^{7-}+\left[\mathrm{CuO}_{4}\right]^{5-},
\end{equation}
instead of
\begin{equation}
	3 d^{9}+3 d^{9} \rightarrow 3 d^{10}+3 d^{8} .
\end{equation}
In any case, ``symmetric'' $d-d$-disproportionation as opposed to ``asymmetric'' ``single-center'' $d-p$-disproportionation [4] has a two-center nature, though can include $d-p$ transfer between the clusters.

Bound electron and hole centers form an electron/hole (EH) dimer characterized by strong coupling with the specific vibrational mode of expansion/compression of the adjacent clusters (half-breathing, or breathing mode). Note that the metastable EH dimers can be also formed as a result of condensation of ``Mott-Hubbard'' $d-d$-excitons with charge transfer.

It can be easily seen that the best conditions for boson superconductivity induced by disproportionation are expected for the parent $3 d^{n}$ systems with transfer of $e_{g}$-electron/hole and with $3 d^{n \pm 1}$-configurations corresponding to maximally stable empty, filled or half-filled $t_{2 q}$ and $e_{g}$-shells. Local composite bosons with $e_{g}^{2}:{ }^{3} A_{2 g}$ configuration with orbital non-degenerate ground state, i.e. $S$-type state, moving in the lattice composed of centers also with $S$-type ground state can actually form only in this case, which allows to minimize the vibronic reduction effect and avoid localization. These are $e_{g}$-electrons/holes that can ensure the maximal magnitude
of the boson transfer integral due to the involvement of the strongest cation-anion $\sigma$-bonds. The transfer integral value depends on the cation-anion bond covalence parameters. Thus, the systems which are the best for superconductivity associated with disproportionation shall be sought for among, for example, oxides, rather than fluorides, because $d-p$-covalence is much stronger in oxides than in fluorides. And finally, in full accordance with Hirsch's concept of hole superconductivity [5,6], hole composite bosons shall be considered as the best ones for HTSC.

The best composite boson configurations and spin as well as orbital state and local spin of the lattice formed as a result of anti-JT-disproportionation in JT magnets with $3 d^{n}$ configuration as well as some $4 d$ JT configurations are listed in the Table I. The last column of the Table I shows the examples of popular $3 d$ - and $4 d$-compounds, whose properties eventually support the conclusions of our model. Among all systems with the possibility of boson superconductivity induced by anti-JT-disproportionation as listed in the Table I, the bulk superconductivity has been reliably observed only in 2D cuprates/nickelates, Fe-pnictides/chalcogenides and ruthenates $\mathrm{Sr}_{2} \mathrm{RuO}_{4}$ and $\mathrm{RuO}_{2}[7,8]$. Multiple experimental and theoretical studies $[9-15]$ show that $\mathrm{LaMnO}_{3}$ and substituted compounds exhibit the properties typical for manifestation of the local spin-triplet superconductivity. Disproportionation has been firmly ascertained in ferrates $(\mathrm{Ca}, \mathrm{Sr}) \mathrm{FeO}_{3}, 3 \mathrm{D}$ nickelate $\mathrm{RNiO}_{3}$ and silver oxide $\mathrm{AgO}$.

Focus shall be made on the predicted spin-triplet superconductivity as a result of anti-JT-disproportionation only for JT magnets with highly symmetric (octahedral, tetrahedral), though distorted, environment. Thereby, a special position of systems with $d^{9}$ configuration, primarily cuprates, in the Table I shall be noted.

$\mathrm{Cu}^{2+}$ in octahedral complexes is characterized by the strongest JT bond and is the most popular almost ,textbook" illustration of the Jahn-Teller effect. This effect results in formation of insulating state of quantum antiferromagnet, for example, in $\mathrm{KCuF}_{3}$ and $\mathrm{La}_{2} \mathrm{CuO}_{4}$. However, as opposed to fluoride, JT distortion in $\mathrm{La}_{2} \mathrm{CuO}_{4}$ results in formation of $\mathrm{CuO}_{2}$-planes with ,perovskite" configuration of $\mathrm{CuO}_{4}$-clusters with the ground $b_{1 \mathrm{~g}} \propto d_{x^{2}-y^{2}}$ state of $e_{g}$-hole, which ensures the strong $\sigma$-bond channel for hole transfer in $\mathrm{CuO}_{2}$ plane and for disproportionation (2) with formation of spin-singlet and orbital non-degenerate $\left({ }^{1} A_{1 g}\right)$ electron center $\left[\mathrm{CuO}_{4}\right]^{7-}$ (equivalent of $\left.\mathrm{Cu}^{+}\right)$ and Zhang-Rice (ZR) hole center $\left[\mathrm{CuO}_{4}\right]^{5-}$ (equivalent of $\mathrm{Cu}^{3+}$ ).

Analysis of multiple experimental data shows that the system of $\mathrm{CuO}_{4}$-centers in $\mathrm{CuO}_{2}$ planes formed either due to the JT effect both in $\mathrm{La}_{2} \mathrm{CuO}_{4}$ and other cuprates with the $T$-structure or due to specific crystal chemistry like in ``apexless'' cuprates with $T^{\prime}$-structure is unstable with respect to charge transfer and disproportionation. Within the charge triplet model [16-22], full disproportionation in $\mathrm{CuO}_{2}$ plane results in formation of the hole and electron center system whose Hamiltonian is equivalent to the Hamiltonian of the system of effective hole spin-singlet composite bosons on non-magnetic lattice formed by electron centers $\left[\mathrm{CuO}_{4}\right]^{7-}$. Such bosons are not conventional quasiparticles, but are an indivisible part of the hole ZR-center.

The absence of fundamental qualitative differences in the electron structure of ``apexless'' nickelates $\mathrm{RNiO}_{2}$ and cuprates, primarily cuprates with $T^{\prime}$-structure, is pointed out in [21,22]. The unusual properties of cuprates and nickelates are the result of ``competition'' of various parameters that govern the ground state of $\mathrm{CuO}_{2}\left(\mathrm{NiO}_{2}\right)$ planes. Thus, while for the vast majority of parent cuprates there is an antiferromagnetic insulating phase, which corresponds to the limit of strong local correlations, this phase was not found in the parent nickelates $\mathrm{RNiO}_{2}$, which can be associated with a lower value or even change of the local correlation parameter sign. In [21,22], we offer that a ,parent" system shall mean cuprate or nickelate with hole half-filling of $\mathrm{CuO}_{4}\left(\mathrm{NiO}_{4}\right)$ centers, which, depending on the parameters of local and non-local correlations, transfer and exchange integrals, as well as ``external'' crystal field formed by the outof-plane environment, can have a different ground state from antiferromagnetic insulator (AFMI), unusual Bose superconductor (BS), Fermi metal (FL) to non-magnetic charge-ordered $(\mathrm{CO})$ insulator. Obviously, these phases will differ not only in electronic, but also in lattice degrees of freedom whose interaction ensures the minimum of the total free energy. In addition, the competition of several possible phases with close energies will lead to phase separation, which will have a significant effect on the observed physical properties. In particular, a so-called pseudogap phase of cuprates is a result of AFMI-CO-BS-FL phase separation [21].

The anti-JT-disproportionation model predicts the possibility of "silver way" to superconductivity in the systems based on $\mathrm{Ag}^{2+}\left(4 d^{9}\right)$, i.e. $4 d$-equivalent of $\mathrm{Cu}^{2+}$. The most probable candidate is silver fluoride $\mathrm{AgF}_{2}$ [23-25] also known as $\alpha-\mathrm{AgF}_{2}$, a perfect equivalent of cuprate $\mathrm{La}_{2} \mathrm{CuO}_{4}$ with surprisingly close electron parameters, but with higher buckling of $\mathrm{AgF}_{2}$ planes. However, this fluoride is a canted antiferromagnetic insulator, though close to instability with charge transfer. Experimental studies [26] report that a metastable disproportionated diamagnetic $\beta-\mathrm{AgF}_{2}$ phase was detected and interpreted as charge-ordered $\mathrm{Ag}^{1+} \mathrm{Ag}^{3+} \mathrm{F}_{4}$ compound, which quickly transforms into $\alpha-\mathrm{AgF}_{2}$ structure.

As opposed to antiferromagnetic insulator $\mathrm{Cu}^{2+} \mathrm{O}$, its silver $4 d$-equivalent $\mathrm{Ag}^{2+} \mathrm{O}$ is a diamagnetic semiconductor with disproportionated $\mathrm{Ag}$ sublattice whose chemical formula is often written as $\mathrm{Ag}^{1+} \mathrm{Ag}^{3+} \mathrm{O}_{2}$ with collinear $\mathrm{O}-\mathrm{Ag}^{1+}\left(4 d^{10}\right)$ - $\mathrm{O}$ bonds and square-planar $\mathrm{Ag}^{3+}\left(4 d^{8}\right) \mathrm{O}_{4}$ bonds $[27,28]$. Moreover, $\left[\mathrm{AgO}_{4}\right]^{5-}$ cluster like $\left[\mathrm{CuO}_{4}\right]^{5-}$ center in cuprates is in non-magnetic state of Zhang-Rice singlet type.

\section{Phase states of Jahn-Teller magnets unstable with respect of charge transfer}

Many experimental data acquired during a long-term study of various properties of a wide class of $2 \mathrm{D}$ cuprates and nickelates and the theoretical simulation of phase diagrams in the charge triplet model [21] give important information regarding possible phase states of JT magnets unstable to charge transfer.

Some long-range order in JT magnets starts forming at high temperatures in a disordered phase for which competition of electron-vibrational interaction, spin and charge fluctuations for low-temperature ground state is typical. Local JT interaction results in stabilization of low-symmetry insulating magnetic structures. Low-energy charge fluctuations of the local anti-JT-disproportionation reaction type (1), depending on the relationship between the local and non-local correlations, one- and two-particle transfer integrals and electron-vibrational interaction parameters with the mode typical for the electron-hole pairs (breathing or half-breathing), can result in formation of a wide range of phases from charge $(\mathrm{CO})$ and spin charge ordering, collinear and noncollinear magnetic ordering, coherent metallic Fermi liquid phase, boson superconductivity phase (BS), and specific quantum EH dimer phase $[21,29]$.

Taking into account the existence of one- or two-particle transport, the high temperature phase for such systems will look like a kind of ``boson-fermion soup'' [30] or ``strange'' metal with $T$-linear resistance dependence and violation of the Ioffe-Regel criterion. Actually, the ``strange'' metal phase is typical for all materials included in the Table I.

Anti-Jahn-Teller disproportionation in ``two-band'' systems of high-spin octa-centers with $3 d^{4}, 4 d^{4}$ configuration or tetrahedral JT centers with $3 d^{6}, 4 d^{6}$ configuration predicts unusual phases with the existence of a system of delocalized effective spin-triplet bosons with $e_{g}^{2}:{ }^{3} A_{2 g}$ configuration moving in the magnetic lattice with localized spin $3 / 2$ and configuration $t_{2 g}^{3}$ (see the Table I), while this does not preclude the existence of unusual phases with localized spin-triplet boson and delocalized $t_{2 g}$-electrons (see the review [31]).

However, which seems more surprising, our simple model provides convincing predictions as to superconductivity and its features in quasi two-dimensional Fepnictides/chalcogenides and ruthenates $\mathrm{Sr}_{2} \mathrm{RuO}_{4}$ and $\mathrm{RuO}_{2}$ which differ in electron structure of active centers and, in particular, in case of $\mathrm{FePn}$ and $\mathrm{FeCh}$, in the local crystalline structure. In both cases, the model predicts hole-type boson spin-triplet superconductivity in $\mathrm{FePn} / \mathrm{FeCh}$ with rather high $T_{c}$ and electron spin-triplet superconductivity in $\mathrm{Sr}_{2} \mathrm{RuO}_{4}$ with very low $T_{C}$, actually in accordance with Hirsch's ideas on the hole nature of HTSC [5,6]. Our model actually assumes that the superconducting carriers in $\mathrm{FePn} / \mathrm{Ch}$ compounds consist of $e_{g}$-holes, rather than of $t_{2 g}$-electrons as predicted by the one-electron multi-orbital band model [32].
Spin-triplet nature of superconducting carriers in $\mathrm{FePn} / \mathrm{FeCh}$ was offered as early as in 2008 [3,33] and supported by a set of experimental data [34-36], though the experimental data is controversial $[37,38]$. It is worth noting that the main current method for determining superconducting carrier spin is the spin susceptibility measurement by Knight shift measurement [31]. It is considered that spins in the triplet superconductor shall be polarized in the external magnetic field like free spins in a conventional metal. Thus, it can be expected in such system that the spin susceptibility and Knight shift shall not have any features in $T_{c}$. Spin anisotropy may suppress this for some directions, but not for others. In a spin-singlet superconductor, magnetic susceptibility vanishes at $T \rightarrow 0$. Thus, for spin-singlet superconductivity, decrease in homogeneous spin susceptibility below $T_{c}$ may be expected, though the same can occur for certain triplet components, though the vanishing susceptibility is often hard to identify due to the background Van Vleck contribution. However, such procedure does not include a complex nature of spin interactions and spin structure of spin-triplet superconductor.

The superconducting state as one of the possible ground states of JT magnets can compete with a normal Fermi liquid state, charge order, spin charge density wave, collinear or noncollinear magnetic order as well as specific quantum phases. Variety of competing phases definitively indicates an important role of phase separation effects $[21,39]$, which shall be considered first of all during the experimental data review.

\section{Effective Hamiltonian of the system of spin-triplet composite bosons}

At variance with the spin-singlet bosons, the effective spin-Hamiltonian of the spin-triplet composite boson system contains a set of additional terms, including generally a conventional magneto-dipole interaction $V_{m d}$, anisotropic boson-boson bilinear and biquadratic exchange interaction, second order single-ion spin anisotropy. However, for the system of $s=1$ bosons in the magnetic lattice, additional terms of local and nonlocal spin-spin interaction between bosons and magnetic lattice occur in the effective spinHamiltonian. More specifically, to describe disproportionated systems, we should consider the electron-lattice interaction primarily with so-called ,half-breathing" mode, but below the effective Hamiltonian of spin-triplet bosons will be discussed in the approximation of a "frozen" lattice.

\subsection{Non-magnetic lattice}

As shown in the Table I, anti-Jahn-Teller disproportionation in the system of tetrahedral JT centers with $3 d^{1}, 4 d^{1}$ configurations, low-spin octa-centers with $3 d^{7}, 4 d^{7}$ configuration or octa-centers with $3 d^{9}, 4 d^{9}$ configuration results in formation of ahalf-filled system of effective spin-triplet
bosons moving in the non-magnetic lattice. The Hamiltonian of such system may be written as
\begin{equation}
	\begin{aligned}
	\mathscr{H}= & -\sum_{i>j, v} t_{i j}\left(\widehat{B}_{i v}^{\dagger} \widehat{B}_{j v}+\widehat{B}_{i v} \widehat{B}_{j v}^{\dagger}\right) \\
	& +V \sum_{i>j, v, v^{\prime}} n_{i v} n_{j v^{\prime}}-\sum_{i, v} \mu_{v} n_{i v}+\mathscr{H}_{s},
	\end{aligned}
\end{equation}
where for composite boson creation/annihilation operators $\widehat{B}_{i v}^{\dagger} / \widehat{B}_{i v}$, independently of spin component $v=0, \pm 1$, Fermi permutation anticommutation relations are satisfied on one site and Bose commutation relations are satisfied for different sites
\begin{equation}
	\left\{\widehat{B}_{i}, B_{i}^{\dagger}\right\}=1, \quad\left[\widehat{B}_{i}, \widehat{B}_{j}^{\dagger}\right]=1
\end{equation}

Fermi anticommutation relations may be rewritten in the form of
\begin{equation}
	\left[\widehat{B}_{i}, \widehat{B}_{i}^{\dagger}\right]=1-2 \widehat{B}_{i}^{\dagger} \widehat{B}_{i}=1-2 \hat{N}_{i}
\end{equation}
In general, these relations preclude double population of a site with bosons. For the spin-independent boson transfer integral in external magnetic field, the standard peierls substitution may be used
\begin{equation}
	t_{i j} \rightarrow t_{i j} e^{i\left(\Phi_{j}-\Phi_{i}\right)}
\end{equation}
wherein
\begin{equation}
	\left(\Phi_{j}-\Phi_{i}\right)=-\frac{q}{\hbar c} \int_{R_{i}}^{R_{j}} \mathbf{A}(r) d \mathbf{l}
\end{equation}
where $\mathbf{A}$ is the vector potential of a homogeneous magnetic field, integration occurs on the line connecting sites $i$ and $j$. The second term in (4) describes inter-site correlations $(V)$, which are similar for different projections of boson spin. Chemical potential $\mu$ is introduced to fix boson concentration $n=\frac{1}{N} \sum_{i \nu}\left\langle\hat{n}_{i \nu}\right\rangle$. Spin-Hamiltonian $\mathscr{H}_{s}$ of the spin-triplet boson system is written as
\begin{equation}
	\begin{aligned}
	& \mathscr{H}_{s}=V_{\mathrm{md}}+\sum_{i>j} J_{i j}\left(\hat{\mathbf{s}}_{i} \cdot \hat{\mathbf{s}}_{j}\right)+\sum_{i>j} j_{i j}\left(\hat{\mathbf{s}}_{i} \cdot \hat{\mathbf{s}}_{j}\right)^{2} \\
	& +K \sum_{i}\left(\mathbf{m}_{i} \cdot \hat{\mathbf{s}}_{i}\right)\left(\mathbf{n}_{i} \cdot \hat{\mathbf{s}}_{i}\right)-\sum_{i}\left(\mathbf{h} \cdot \hat{\mathbf{s}}_{i}\right) \ldots,
	\end{aligned}
\end{equation}
where only few typical terms are highlighted, $J_{i j}$ and $j_{i j}$ are bilinear and biquadratic isotropic exchange integrals, respectively, $K$ is the constant, and $\mathbf{m}$ and $\mathbf{n}$ are the unit vectors generally defining two typical axis of the second order single-ion anisotropy, $\mathbf{h}$ is the external field.

\subsection{Magnetic lattice and double exchange}

However, the anti-Jahn-Teller disproportionation in the system of high-spin octa-centers with $3 d^{4}, 4 d^{4}$ configuration or tetrahedral JT-centers with $3 d^{6}, 4 d^{6}$ configuration results in formation of a system of effective spin-triplet bosons with $e_{g}^{2}:{ }^{3} A_{2 g}$ configuration moving in the magnetic lattice with localized spins $S=3 / 2$ with $t_{2 g}^{3}$ configurations (see the Table I). The effective Hamiltonian of such system may be also written as (4), but the spin-Hamiltonian $\mathscr{H}$ will have a much more complicated structure. Taking into account only bilinear spin-spin isotropic exchange, this may be written as

\begin{equation}
	\begin{aligned}
	\mathscr{H}_{s}= & \sum_{i>j} J_{i j}^{l l}\left(\widehat{\mathbf{S}}_{i} \cdot \widehat{\mathbf{S}}_{j}\right)+\sum_{i>j} J_{i j}^{b b}\left(\hat{\mathbf{s}}_{i} \cdot \hat{\mathbf{s}}_{j}\right) \\
	& +\sum_{i \neq j} J_{i j}^{b l}\left(\hat{\mathbf{s}}_{i} \cdot \widehat{\mathbf{S}}_{j}\right)+\sum_{i} J_{i i}^{b l}\left(\hat{\mathbf{s}}_{i} \cdot \widehat{\mathbf{S}}_{i}\right)
	\end{aligned}
\end{equation}
where the first term describes the ``lattice'' spin exchange, the second term describes the exchange interaction between spin-triplet bosons, the third and fourth term describe the exchange between bosons and lattice spins, while the latter term actually describes the Hund intraatomic exchange. In order for Hund's rule to be fulfilled, the exchange integral $J_{i i}^{b l}$ shall be assumed as a large ferromagnetic integral. Taking into account transfer of spin-triplet bosons, we actually arrive at a Bose analogue of the simplest model of double exchange [9].

\section{Conclusion}

Unusual properties of a wide class of materials based on Jahn-Teller $3 d$ - and $4 d$-ions with different crystalline and electronic structures, from quasi two-dimensional unconventional superconductors (cuprates, nickelates, Fe-pnictides/chalcogenides, ruthenate $\mathrm{Sr}_{2} \mathrm{RuO}_{4}$ ), manganites with local superconductivity to $3 \mathrm{D}$ ferrates $(\mathrm{CaSr}) \mathrm{FeO}_{3}$, nickelates $\mathrm{RNiO}_{3}$ and $\mathrm{AgO}$ with unusual charge and magnetic order, may be explained within a single scenario where their instability with respect to the anti-Jahn-Teller disproportionation is assumed. These systems feature formation of effective local composite spin-singlet or spintriplet, electron or hole $S$-type bosons moving in nonmagnetic or magnetic lattice, which results in extremely wide range of phase states from non-magnetic and magnetic insulators to unusual metallic and superconducting states. The anti-JT disproportionation model predicts the spintriplet superconductivity in ruthenates $\mathrm{Sr}_{2} \mathrm{RuO}_{4}$ and $\mathrm{RuO}_{2}$, Fe-pnictides/chalcogenides $\mathrm{FePn} / \mathrm{FeCh}$, manganite $\mathrm{LaMnO}_{3}$, though one or another spin-charge order is implemented in well-known ,candidates" $\left(\mathrm{Ca}(\mathrm{Sr}) \mathrm{FeO}_{3}, \mathrm{RNiO}_{3}, \mathrm{AgO}\right)$. The model assumes that the superconducting carriers in $\mathrm{FePn} / \mathrm{Ch}$ compounds consist of $e_{g}$-holes, rather than of $t_{2 g}$-electrons as predicted by the one-electron multi-orbital band model. The best conditions for HTSC with spinless local bosons and spinless lattice can be achieved only for low-symmetry quasi two-dimensional $d^{9}$-systems such as 2D cuprates and nickelates. Effective Hamiltonians for spin-triplet composite bosons in non-magnetic and magnetic lattices have a complicated spin structure, which shall be taken into account for interpretation of superconducting carrier spin experiments.

\section{Funding}

This study was supported by project FEUZ-2023-0017 of the Ministry of Science and Higher Education of the Russian Federation.

\section{Conflict of interest}

The authors declare that they have no conflict of interest.

\section*{References}
{
\setlength\parindent{0pt}
\setlength{\parskip}{6pt}

[1] A.J. Leggett, Y. Liu. J. Supercond. Nov. Magn. \textbf{34}, 1647 (2021).

[2] A.S. Moskvin. J. Phys.: Condens. Matter 25, 085601 (2013).

[3] A.S. Moskvin, I.L. Avvakumov. Proc. III Int. Conf. ``Fundamental Problems of High-Temperature Superconductivity'' (Moscow, Zvenigorod, 13-17 October 2008) p. 215.

[4] S. Mazumdar. Phys. Rev. B \textbf{98}, 205153 (2018). [Phys. Rev. Res. 2, 023382 (2020)]

[5] J.E. Hirsch. Proc. SPIE 10105. Oxide-based Materials and Devices VIII, 101051V (7 March 2017).

[6] J.E. Hirsch, F. Marsiglio. Physica C \textbf{564}, 29 (2019).

[7] J.P. Ruf, H. Paik, N.J. Schreiber, H.P. Nair, L. Miao, J.K. Kawasaki, J.N. Nelson, B.D. Faeth, Y. Lee, B.H. Goodge, B. Pamuk, C.J. Fennie, L.F. Kourkoutis, D.G. Schlom, K.M. Shen. Nature Commun. \textbf{12}, 59 (2021).

[8] M. Uchida, T. Nomoto, M. Musashi, R. Arita, M. Kawasaki. Phys. Rev. Lett. \textbf{125}, 147001 (2020).

[9] A.S. Moskvin. Phys. Rev. B \textbf{79}, 115102 (2009).

[10] Kim Yong-Jihn. Mod. Phys. Lett. B \textbf{12}, 507 (1998).

[11] V.N. Krivoruchko. Low Temp. Phys. \textbf{47}, 901 (2021).

[12] V. Markovich, I. Fita, A. Wisniewski, R. Puzniak, D. Mogilyansky, L. Titelman, L. Vradman, M. Herskowitz, G. Gorodetsky. Phys. Rev. B \textbf{77}, 014423 (2008).

[13] M. Kasai, T. Ohno, Y. Kauke, Y. Kozono, M. Hanazono, Y. Sugita. Jpn. J. Appl. Phys. \textbf{29}, L2219 (1990).

[14] A.V. Mitin, G.M. Kuz'micheva, S.I. Novikova. Russ. J. Inorg. Chem. \textbf{42}, 1791 (1997).

[15] R. Nath, A.K. Raychaudhuri, Ya.M. Mukovskii, P. Mondal, D. Bhattacharya, P. Mandal. J. Phys. Condens. Matter \textbf{25}, 15, 155605 (2013).

[16] A.S. Moskvin. Phys. Rev. B \textbf{84}, 075116 (2011).

[17] A.S. Moskvin, Y.D. Panov. J. Supercond. Nov. Magn. \textbf{32}, 61 (2019).

[18] A.S. Moskvin, Yu.D. Panov. Physics of the Solid State \textbf{61}, 1553 (2019).

[19] A.S. Moskvin. Phys. Met. Metallogr. \textbf{120}, 1252 (2019).

[20] A. Moskvin, Y. Panov. Condens. Matter \textbf{6}, 24 (2021).

[21] A.S. Moskvin, Yu.D. Panov. JMMM \textbf{550}, 169004 (2022).

[22] A.S. Moskvin. Optics and Spectroscopy \textbf{131}, 491 (2023).

[23] P. Fischer, G. Roult, D. Schwarzenbach. J. Phys. Chem. Solids \textbf{32}, 1641 (1971).

[24] M. Derzsi, K. Tokar, P. Piekarz, W. Grochala. Phys. Rev. B \textbf{105}, L081113 (2022).

[25] N. Bachar, K. Koteras, J. Gawraczynski, W. Trzciński, J. Paszula, R. Piombo, P. Barone, Z. Mazej, G. Ghiringhelli, A. Nag, Ke-Jin Zhou, J. Lorenzana, D. van der Marel, W. Grochala. Phys. Rev. Res. \textbf{4}, 023108 (2022).

[26] C. Shen, B. Zemva, G.M. Lucier, O. Graudejus, J.A. Allman, N. Bartlett. Inorg. Chem. \textbf{38}, 4570 (1999).
[27] V. Scatturin, P.L. Bellon, A.J. Salkind. J. Electrochem. Soc. \textbf{108}, 819 (1961).

[28] J.P. Allen, D.O. Scanlon, G.W. Watson. Phys. Rev. B \textbf{84}, 115141 (2011).

[29] A.S. Moskvin, Yu.D. Panov. Physics of the Solid State \textbf{62}, $1554(2020)$.

[30] E. Pangburn, A. Banerjee, H. Freire, C. Pepin. Phys. Rev. B \textbf{107}, 245109 (2023).

[31] P.J. Hirschfeld. Comptes Rendus Phys. \textbf{17}, 197 (2016).

[32] G.R. Stewart. Rev. Mod. Phys. \textbf{83}, 1589 (2011).

[33] P.A. Lee, Xiao-Gang Wen. Phys. Rev. B \textbf{78}, 144517 (2008).

[34] S.-H. Baek, H.-J. Grafe, F. Hammerath, M. Fuchs, C. Rudisch, L. Harnagea, S. Aswartham, S. Wurmehl, J. van den Brink, B. Büchner. Eur. Phys. J. B \textbf{85}, 159 (2012).

[35] T. Hanke, S. Sykora, R. Schlegel, D. Baumann, L. Harnagea, S. Wurmehl, M. Daghofer, B. Büchner, J. van den Brink, C. Hess. Phys. Rev. Lett. \textbf{108}, 127001 (2012).

[36] P.M.R. Brydon, M. Daghofer, C. Timm. J. van den Brink. Phys. Rev. B \textbf{83}, 060501(R) (2011).

[37] J. Brand, A. Stunault, S. Wurmehl, L. Harnagea, B. Büchner, M. Meven, M. Braden. Phys. Rev. B \textbf{89}, 045141 (2014).

[38] J.A. Gifford, B.B. Chen, J. Zhang, G.J. Zhao, D.R. Kim, B.C. Li, D. Wu, T.Y. Chen. AIP Adv. \textbf{6}, 115023 (2016).

[39] A.S. Moskvin, Yu.D. Panov. J. Phys.: Conf. Ser. \textbf{2164}, 012014 (2022).
}

\end{document}